# Near infra-red optical materials from polymeric amorphous carbon synthesised by collisional plasma process

M. Rybachuk [1,2], A. Hertwig [1], M. Weise [1], M. Sahre [1], M. Männ [1], U. Beck [1] and J. M. Bell [2]

[1] *Federal Institute for Materials Research and Testing (BAM), Division VI.4 Surface Technology, Unter den Eichen 87, 12205 Berlin, Germany*

[2] *Faculty of Built Environment and Engineering, Queensland University of Technology, 2 George St, Brisbane, Qld 4001, Australia*

ABSTRACT

The synthesis of polymer-like amorphous carbon ($a$-C:H) thin-films by microwave excited collisional hydrocarbon plasma process is reported. Stable and highly aromatic $a$-C:H were obtained containing significant inclusions of poly($p$-phenylene vinylene) (PPV). PPV confers universal optoelectronic properties to the synthesised material. That is $a$-C:H with tailor-made refractive index are capable of becoming absorption-free in visible (red) – near infrared wavelength range. Production of large aromatic hydrocarbon including phenyl clusters and/or particles is attributed to enhanced coagulation of elemental plasma species under collisional plasma conditions. Detailed structural and morphological changes that occur in $a$-C:H during the plasma synthesis are also described.

MAIN TEXT

The applications of amorphous diamond-like carbon (DLC) and/or hydrogenated carbon (*a*-C:H) materials as photonic materials demand both, the passivation of dangling bonds by hydrogen and a high degree of aromaticity of $sp^2$-phase (*e.g.* aryl and/or heteroarene hydrocarbon compounds) as those equally contribute to minimising optical losses. Hydrocarbon plasma methods used for the synthesis of *a*-C:H provide precise control over thin-film surface finish, thin-film thickness and uniformity of phase-composition of the coating but fail in producing *a*-C:H with satisfactory optoelectronic properties. This owns to the intrinsic thin-film nucleation process that is governed by complex kinetic reactions of radical species (*CH$_3$* mono-radicals, di-radicals, and/or *C$_2$H$_4$* and *C$_2$H$_2$* un-saturated species) in plasma.[1,2] These radicals can form large aromatic compounds (*i.e.* stable intermediates) following multiple gas-phase nucleation reactions and polymerize with other radicals and/or hydrocarbon monomers. For these radicals to agglomerate into aryl compounds extended (*i.e.* long) gas-phase propagation pathways are required. The latter, we suggest, could be attained with the mean-free-path being less than the physical path length to the substrate under common *a*-C:H synthesis experiments.

In this letter, we report a simple method that allows the fabrication of highly aromatic polymer-like *a*-C:H materials using a microwave excited hydrocarbon plasma operated in collisional mode. *a*-C:H were found to contain large inclusions of poly(*p*-phenylene vinylene) (PPV) which awards characteristic optoelectronic properties that have not been reported previously. That is, *a*-C:H thin-films could become absorption-free in visible (red) – near infrared (N-IR) wavelength range (600 – 1200 nm) with their refractive index tailored. Our results exemplify the formation of phenyl clustering of $sp^2$-phase in $a$-C:H and support the theoretical simulations of Bleecker et al.[3] for low-pressure acetylene discharges with stable intermediate aromatic hydrocarbon formation. Furthermore, our work demonstrates the potential for the use of collisional plasma process for fabrication of materials with previously unattainable composition and properties, and postulate this is not limited to DLC, $a$-C:H or other carbonaceous materials.

$a$-C:H films were deposited on *Si* <111> using a $C_2H_2$/*Ar* gas mixture in a modified 2.45 GHz microwave excited electron-cyclotron resonance (MW-ECR) plasma reactor[4] (Von Ardenne Anlagentechnik GmbH/Roth & Rau AG) at temperatures of ~300 K and pressure of ~0.1 Pa. In MW-ECR system the physical path length to the substrate was increased one order of magnitude relative to the mean-free-path from ~$1.5 \times 10^{-2}$ m to ~$2.9 \times 10^{-1}$ m while retaining the original MW-ECR power density characteristics. The mean-free-path (for 300 K, 0.1 Pa) was found to be ~$2.9 \times 10^{-2}$ m for hydrocarbon molecules measuring ≤ 4 Å. Experimentally the plasma power was varied from 50 W to 1200 W in 200 W increments.

The $a$-C:H $sp^2$-phase was examined *ex situ* at 293 K using unpolarized Raman light of 532 nm (Renishaw) recorded in a backscattering geometry with precautions taken to avoid sample damage.[5] The amount of hydrogen, $H$, (at. %) was estimated using basic Raman parameters for the $D$ and $G$ peaks and the slope of the photoluminescence (PL) background from a spectrum taken at 532 nm following the work of Casiraghi et al.[6] Analysis of $C_{1s}$ core-valence bands of X-ray photoelectron spectra (Kratos Axis Ultra) determined $sp^3$-phase content in the films.[7] Film thickness, band gap, $E_g$, and optical constants (refractive index, $n$, and extinction coefficient, $k$) were measured in ultraviolet (UV) - N-IR range as a function of wavelength, $\lambda$ (nm), by UV-IR spectroscopic ellipsometry (J.A. Woollam Co., M2000 DI). The $n(\lambda)$ and $k(\lambda)$ spectra were analysed employing a Kramers - Kronig oscillator model and by fitting a single Tauc - Lorentzian absorption constituent.[4,8] Thin film growth rates were found varying from 15 nm/min (at 50 W) to 110 nm/min (at 1200 W). Mass density of $a$-C:H samples was obtained using X-ray reflectometer (Seifert XRD 3000 TT) collected with $Cu\ K_\alpha$ radiation at 40 kV and 40 mA.

A set of Raman spectra collected from the $a$-C:H thin-films at 532 nm is shown in Fig. 1(a). The first order region (900 – 2000 cm$^{-1}$) is due to the overlap of fundamental Raman $D$ and $G$ bands[7,9] indicating the abundance of $sp^2$-phase arranged in both aromatic rings and olefinic chains. Notably, the spectra change from a single asymmetric peak superimposed on a high PL background (*e.g.* at 50 W; $sp^2$-phase is olefinic) to double peaks superimposed on a decreasing PL background ($sp^2$-phase is highly aromatic). A small contribution from a peak positioned in 840 – 870 cm$^{-1}$ region is also seen (see inset) analogous to the 867 cm$^{-1}$ band reported by Arenal *et al.*[10] for highly aromatic and 'near-frictionless' $a$-C:H. Extended crystal modes of large aromatic clusters of $\pi$-bonded rings exhibit excitations in this frequency range. These clusters permit phonon excitations from outside the center of the Brillouin zone (with wave-vector $k \neq 0$) to contribute to Raman scattering due to the breakdown in translational symmetry.[5-7,10] Fig. 1(b) shows a fitted spectrum of $a$-C:H grown at 400 W. The fitted bands (Gaussian lineshapes) are $D$ and $G$ modes, two $Ag$ zone center vibrational modes of *trans*-polyacetylene (*trans*-(CH)$_x$), $\omega_1$ at ~1060 cm$^{-1}$ (*C-C* stretching) and $\omega_3$ at ~1450 cm$^{-1}$ (*C=C* stretching), notably the contributions from the $\omega_1$ and $\omega_3$ bands in the fitting are minimal, and two additional vibrational modes which were found in all but the sample fabricated at 50 W. These are an 1175 cm$^{-1}$ mode and a mode with a frequency in 1530 – 1545 cm$^{-1}$ range. Recently, an *a*-C:H

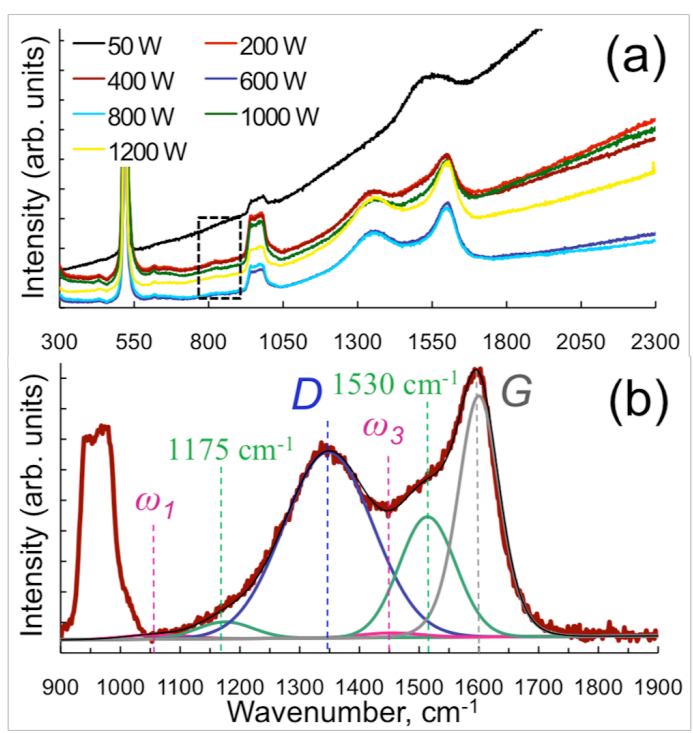

FIG. 1 532 nm Raman spectra of *a*-C:H; (a) Non-normalized spectra of *a*-C:H synthesized under increasing microwave power showing decreasing PL background, and (b) Fitted spectrum of an *a*-C:H (400 W) showing contributions from *D* and *G* modes, *trans*-(*CH*)$_x$, $\omega_1$ and $\omega_3$ modes, and PPV 1175 cm$^{-1}$ and 1530 cm$^{-1}$ modes. An asymmetric peak at 950 cm$^{-1}$ is the second order *Si*.

Raman peak at 1175 cm$^{-1}$ has been identified as combination of a vinylene and a *CC–H* ring bend modes in neutral PPV.[5,7] Other PPV zone centre vibrational modes are found at higher frequencies in 1520 – 1540 cm$^{-1}$ and 1585 – 1610 cm$^{-1}$ ranges.[11,12] In

*a*-C:H these modes have neither been visible nor reported as they become consistently obscured by the host native *D* and *G* modes. Also, excitations at ~1530 cm$^{-1}$ are certainly not of *trans*-(CH)$_x$ $\omega_3$ mode as the $\omega_3$ peak is positioned at a lower excitation frequency of ~1450 cm$^{-1}$ when probed by green light laser.[7,13] We, therefore, postulate that the origin of a peak observed at ~1530 cm$^{-1}$ is due to phenyl ring stretching of neutral PPV.

The evolution of $sp^2$-phase is shown in Fig. 2, where the *I*(*D*)/*I*(*G*) ratio, Fig. 2(a), *G* peak position (cm$^{-1}$), Fig. 2(b), full-width-at-half-maximum (FWHM) for *G* peak, $\Gamma_G$, Fig. 2(c), and the total relative intensity for PPV contributions, *I*Σ(PPV), Fig. 2(d) are shown relative to the microwave ionization power. The *I*Σ(PPV) is sum of relative intensities of 1175 cm$^{-1}$, *I*(1175), and 1530 cm$^{-1}$, *I*(1530), peaks. Detailed analysis of *I*(*D*)/*I*(*G*) ratio, *G* peak position and $\Gamma_G$ trends reveal that the degree of aromatic clustering is not significantly altered with changing ionization energy except for 50 W samples that are highly olefinic. The *D* peak position was found at 1350(±1) cm$^{-1}$ for all films with FWHM gradually decreasing from 198 to 168 cm$^{-1}$ as microwave power increases. Since the *D* mode is only related to aromatic rings, the small increase of the ratio *I*(*D*)/*I*(*G*) (see Fig. 2(a)) coupled with the narrowing of the *D* peak width demonstrates a decrease in size of $sp^2$-phase clusters. The overall reduction of PPV contributions (see Fig. 1(c)) support this finding confirming that both 1175 cm$^{-1}$ and 1530 cm$^{-1}$ vibrational modes are of PPV origin. Furthermore, 532 nm laser excitation innately probes the PPV band gap that is ~2.3 eV.[14]

Results for hydrogen content (estimated[6] using values for *I*(*D*)/*I*(*G*), PL slope and the *G* peak position) indicate that increase of microwave ionization energy essentially results in de-hydrogenation of *a*-C:H materials (see Fig. 3(a)). Overall Raman results affirm that topological order of *a*-C:H increases at higher power. This unusual trend is attributed to the collisional plasma process where coagulation of hydrocarbon radicals into large clusters (via chain polymerization reactions) occurs via a decrease of average ionization energy. The dissociation energies for $C–H$ bonds (acetyl, internal carbon, terminal carbon, or aromatic hydrogen) are almost half of those for $C=C$ and $C\equiv C$ bonds (3.9 – 4.5 eV vs. 7 – 10 eV)[2] therefore, the average decrease of ionization energy results in de-hydrogenation of $a$-C:H. Findings by Doyle[15] for acetylene discharges reveal that $a$-C:H growth is dominated by $C_2H$ radicals for small gas depletions but $C_4H_3$ and $C_6H_3$ radicals govern high degrees of precursor gas dissociation described by the Bauer–Aten mechanism for benzene decomposition.[16] The emergence of large saturated radicals and/or neutrals at higher ionisation power accounts for the higher topological order and increased concentration of six-fold clusters (*i.e.* PPV) and shows as a reduction of $sp^3$-phase content, see Fig. 3(b). The most unusual feature of *a*-C:H synthesis by collisional plasma is that hydrogen-depleted *a*-C:H materials display both narrower band gap, $E_g$, (Fig. 3(c)) and, are of lower mass density (see Fig. 3(d)). This deviates from the properties of *a*-C:H materials synthesized by collisionless plasma process, where hydrogen concentration is found to be proportional to both $E_g$ and mass-density.[6,10,17,18] These *a*-C:H materials however, do not contain PPV as in our case. We consider that $\pi$ states undergo gradual localization on aromatic clusters ($sp^2$- phase) as hydrogen is reduced once more supporting the origin of phenyl clustering in *a*-C:H.

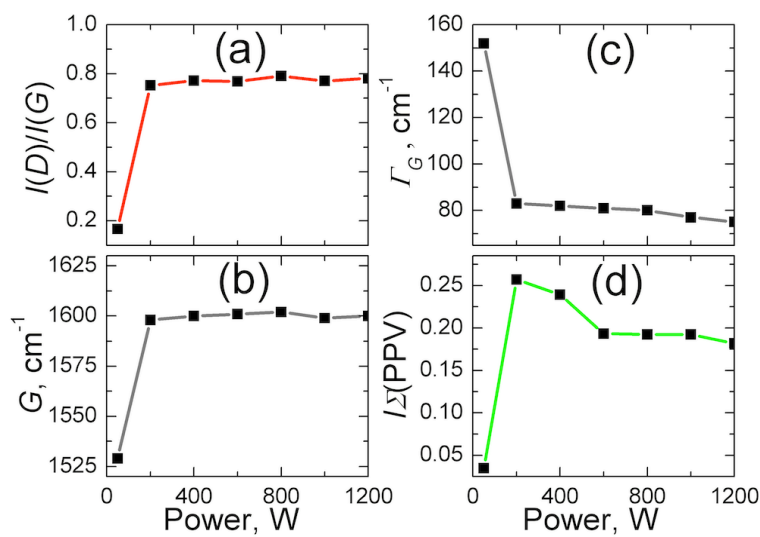

FIG. 2 Variation of (a) $I(D)/I(G)$ ratio, (b) $G$ peak position (cm$^{-1}$), (c) FWHM for $G$ peak and, (d) the total relative intensity for PPV contributions, $I\Sigma$(PPV) as a function of changing ionization power.

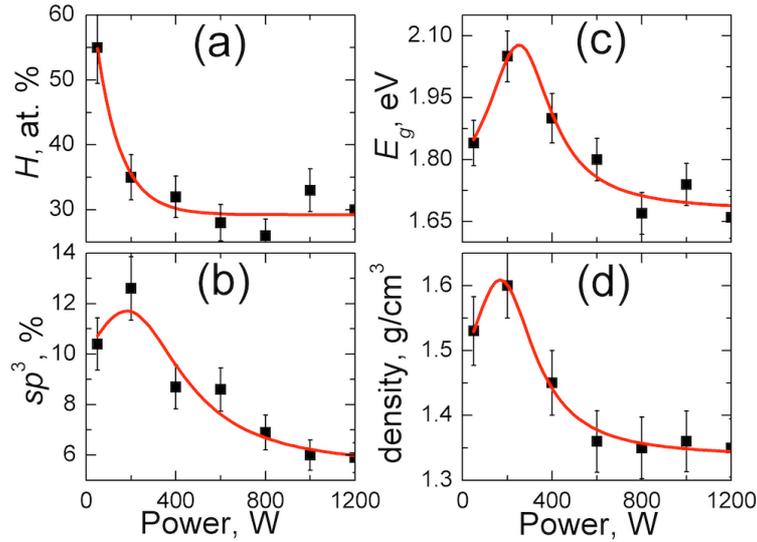

FIG. 3 Variation of (a) hydrogen content, (b) $sp^3$-phase amount, (c) band gap and, (d) mass-density as a function of microwave plasma power. Data points were fitted with an independent variable Lorentzian line-shape.

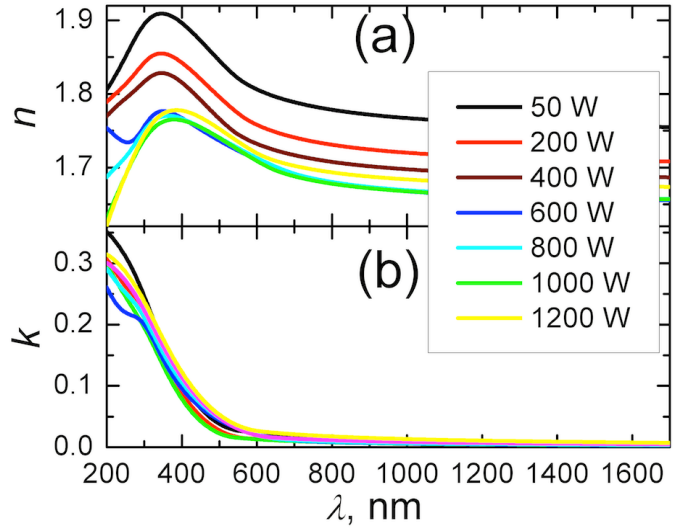

FIG. 4 Variation of (a) *n* and, (b) *k* optical constants for *a*-C:H as a function of microwave ionisation power.

Variation of *a*-C:H optical *n* and *k* constants as a function of wavelength, $\lambda$ under varying microwave power is illustrated in Fig. 4. Fig. 4(a) shows *n* and Fig. 4(b) *k* dependencies. Synthesized *a*-C:H materials exhibit a remarkably low absorption beginning from red light range to and above 1200 nm owing to highly organized $sp^2$-phase. The extinction coefficient was found to be ≤ 0.005 in N-IR region thus making the coatings suitable as dielectric thin layer materials in this wavelength range.

In summary, we have performed the study of *a*-C:H thin-films synthesised by collisional hydrocarbon plasma process under varying microwave plasma ionisation power. It was found that the topological and structural ordering of $sp^2$-phase clusters in *a*-C:H increases with increasing ionisation power. The increase of ionisation, however, leads to simultaneous de-hydrogenation of *a*-C:H. *a*-C:H materials were found to contain large aromatic hydrocarbons, which we deduce are inclusions of PPV (1175 cm$^{-1}$ and 1530 cm$^{-1}$ Raman modes). Formation of PPV in *a*-C:H is attributed to coagulation of acetylene radical species through multiple kinetic chemical reactions (*i.e.* via extended gas-phase propagation pathways) into aromatic clusters and/or particles under collisional plasma conditions. All *a*-C:H displaying significant $sp^2$ and phenyl clustering exhibit a small Raman band in the 840 – 870 cm$^{-1}$ region confirming the observation by Arenal *et al.*[10] Thus, we identify significant structural changes in *a*-C:H with saturation of dangling bonds and cluster coagulation process governed by large, non-propargylic arene species nucleating in the gas-phase. We, therefore, attribute the exceptionally low optical absorption of synthesized *a*-C:H materials in N-IR range to the high aromaticity and presence of phenyl clusters.

ACKNOLEGEMENTS

This work was financially supported by internal funding from QUT and BAM. M.R gratefully acknowledges QUT for postdoctoral support and BAM for visiting research fellowship funding.